\documentclass[conference,final]{IEEEtran}
\pdfoutput=1
\IEEEoverridecommandlockouts
\def\BibTeX{{\rm B\kern-.05em{\sc i\kern-.025em b}\kern-.08em
    T\kern-.1667em\lower.7ex\hbox{E}\kern-.125emX}}

\usepackage{cite}
\usepackage{amsmath,amssymb,amsfonts}
\usepackage{algorithmic}
\usepackage{graphicx}
\usepackage{textcomp}
\usepackage{xcolor}
\usepackage{hyperref}
\usepackage{fancyhdr}
\fancypagestyle{copyright}{%
    \fancyhf{}
    \cfoot{\footnotesize%
        \textcopyright~2021 IEEE. Personal use of this
        material is permitted. Permission from IEEE must be obtained for all other
        uses, in any current or future media, including reprinting\slash republishing
        this material for advertising or promotional purposes, creating new collective
        works, for resale or redistribution to servers or lists, or reuse of any
        copyrighted component of this work in other works. This material is referenced
        by DOI: {\href{https://doi.org/10.1109/BigData52589.2021.9672060}{10.1109/BigData52589.2021.9672060}}
    }
}

\begin{document}
\title{{\it AutoGeoLabel}:\\Automated Label Generation\\for Geospatial Machine Learning}
\author{\IEEEauthorblockN{Conrad M Albrecht}
\IEEEauthorblockA{\textit{Remote Sensing Technology Institute} \\
\textit{German Aerospace Center}\\
We\ss ling, Germany \\
conrad.albrecht@dlr.de}
\and
\IEEEauthorblockN{Fernando Marianno}
\IEEEauthorblockA{\textit{Data Intensive Physical Analytics}\\
\textit{IBM T.J. Watson Research Center}\\
Yorktown Heights, NY, USA \\
fjmarian@us.ibm.com}
\and
\IEEEauthorblockN{Levente J Klein}
\IEEEauthorblockA{\textit{Data Intensive Physical Analytics}\\
\textit{IBM T.J. Watson Research Center}\\
Yorktown Heights, NY, USA \\
kleinl@us.ibm.com}
}

\maketitle
\begin{abstract}
A key challenge of supervised learning is the availability of human-labeled data.
We evaluate a big data processing pipeline to auto-generate labels for
remote sensing data. It is based on rasterized statistical features extracted from
surveys such as e.g. LiDAR measurements. Using simple combinations of the rasterized
statistical layers, it is demonstrated that multiple classes can be generated
at accuracies of $\sim0.9$.\newline
As proof of concept, we utilize the big geo-data platform IBM PAIRS to dynamically
generate such labels in dense urban areas with multiple land cover classes.
The general method proposed here is platform independent, and it can be adapted
to generate labels for other satellite modalities in order to enable machine learning on
overhead imagery for land use classification and object detection.
\end{abstract}
\thispagestyle{copyright}

\section{Introduction \& Motivation}
\label{sec:Intro}

Driven by the availability of massive amounts of data, image classification achieves accuracy
of over 90\% with thousands of classes, today \cite{yalniz}. To a large extent, the performance of modern
deep learning models is driven by the volume of data, and the availability of accurate labels for
training. Popular datasets such as ImageNet \cite{imagenet}, COCO \cite{coco}, and MNIST
\cite{mnist} serve as benchmarks for assessing machine learning model performance.
These benchmarks were generated by visual inspection of images to manually craft labels.
Extension to new labels or classes requires additional effort. The same procedure---as previously executed 
for the collection of existing labels---needs application to the new, unlabeled set of data.

While image classification achieved tremendous success for photographs, a straightforward adoption of
established machine learning techniques for satellites, drones, and remote sensing images have limited
success \cite{penatti,remote-labels,maggiori,wang1}. In general, geospatial image classification
\cite{dlearth} lags behind machine learning precision from social media \cite{mahajan2018}---due
to more sparsely labeled data, and due to increased heterogeneity in data. 

Remote sensing signals often demand far more sophisticated processing compared with conventional
photography. For instance, Light Detection and Ranging (LiDAR) \cite{LiDAR}---acquired as a
3D point cloud---requires processing and conversion of the data into a commonly usable machine
learning 2D data input format; such as a multi-dimensional
array of numbers \cite{paszke2017automatic,tensorflow2015-whitepaper}.
Geospatial data also come with varying spatial resolution acquired across multiple seasons.
Real time integration of satellite data into multidimensional models
like weather or climate requires automation of label generation. 
For example: Identification of extreme weather events such as a hurricane,
earthquakes, wildfires, etc., demands specific training data with need to generate labels on the fly as
those events are forming. 

Here we propose \textit{AutoGeoLabel}, a framework to create labels for geospatial imagery. It
extracts features using simple statistical methods from raw and unlabeled data. The approach allows to identify
class labels from a combination of rasterized layers. We demonstrate the utility of such an approach taking
LiDAR measurements to exemplify how simple combinations of statistical features can help to
automatically classify a geospatial scene. As a result, such rapidly generated labels may get exploited
to identify land cover as (noisy) input for machine learning models. 

The framework is developed to label large volumes of data, and to create labels in near
real time for any image type. Moreover, the method is not limited to LiDAR, but
may get adopted for any other high quality data available for the area of interest such as hyper-
and multi-spectral imaging, synthetic aperture radar (SAR), or microwave radiometry.

\section{Previous Work \& Applications}
\label{sec:PrevWorkAndApp}

Labeled geospatial data for machine learning benchmarks is hard to come by, and those is sensor platform
specific. Commonly, labels from land classification is typically rooted in a combination of
pixel- and object-based classifications \cite{corine}. Generated land cover data is often corrected
through a data quality process where manual reclassification is carried out. Additional maps
and land use datasets may get exploited in the process.
Given the (human) effort involved in generating land cover data,
those is limited in spatial coverage or temporal updates. Typical refresh rates read once in about 3--5 years
\cite{corine, mrlc}. Standard classified data like \textit{CORINE Land Cover} \cite{corine} and
\textit{Multi-Resolution Land Characteristics} (MRLC) \cite{mrlc} products have a spatial resolution of tens of
meters, limited spatial coverage, and the number of classes is restricted to the most common land covers---like forest,
water bodies, and agricultural lands, etc. Standard geospatial benchmark datasets like \textit{Spacenet} \cite{spaenet6},
\textit{BigEarthNet} \cite{BigEarthNetAL}, and \textit{DeepSat}\cite{deepsat} have an even more
limited number of labeled classes.

In contrast, \textit{OpenStreetMap} (OSM) is one of the most complete, crowd-sourced
community efforts with hundreds of labels collected by volunteers \cite{openstreetmap}. OSM labels is
represented in vector shape format (points, lines, polygons) from which rasterized maps can get generated
where roads, houses, buildings, and vegetation-covered areas is color-coded. It has been demonstrated
that such OSM-based land classification labels may train deep neural network models for semantic
segmentation of high-resolution, multi-spectral imagery \cite{albrecht}. In particular, uneven
OSM-label completeness of different geographies added noise to the segmentation task. Technically, semantic
segmentation was handled by unsupervised image-style translation employing a modified CycleGAN
\cite{zhang2020map} architecture.

Concerning remote sensing image classification in general, standard machine learning tools previously
applied read e.g.\ \textit{Random Forest} \cite{randomf}, \textit{Support Vector Machine} \cite{svm},
\textit{XGBoost} \cite{sclassification}, and a plethora of deep learning models \cite{zhu2017deep}.

Geospatial data is one of the most prevalent data in Energy \& Utility, Oil \& Gas,
Forestry, Agriculture, Transportation, Navigation, and Disaster Emergency Response \cite{bolstadgis}.
Many of the above industries do have massive amounts of data collected through conventional observation, sensing, and
various other measurement methods. However, the absence of labels impedes most of automation and predictive methods
based on machine learning.
\begin{figure}
    \centering
    \includegraphics[width=0.95\columnwidth]{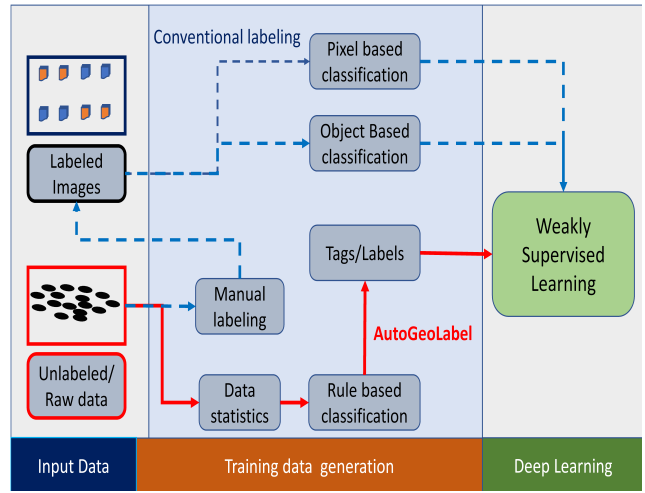}
    \caption{\label{fig:AutoGeoLabelFlowChart}
    Schematic flow chart of the \textit{AutoGeoLabel} framework to generate labeled data for machine learning.
    Motivated by the shortage in geospatial labels (Sec.\ \ref{sec:Intro}) where at the same time there
    exists vast amounts and a plurality of unstructured geospatial information (Sec. \ref{sec:BigGeoDataPlatform}),
    \textit{AutoGeoLabel}  automatically generates
    labels for weakly supervised learning.
    }
\end{figure}
For many industries, the large number of features of interest requires dedicated
efforts to create labeled datasets based on domain expertise.
\textit{AutoGeoLabel} may address such situations where labels and images is
generated on the fly and computational resources is limited. \textit{AutoGeoLabel} bears potential
for lightweight computational techniques ready for deployment on e.g.\ \textit{Internet-of-Things}
(IoT) \cite{madakam2015internet} edge devices \cite{chen2019deep}.

\section{Datasets and Geospatial Platform}
\label{sec:BigGeoDataPlatform}

\subsection{LiDAR Data}
\label{sec:LiDAR}
LiDAR data generate a dense point cloud from laser pulses bouncing back from the Earth's surface.
Massive amounts of LiDAR data is acquired mainly to map topography.
In addition to elevation mapping, LiDAR carries rich embedded information on land cover such as
vegetation, roads, buildings, etc.
However, extracting such features may require reclassification of the point cloud, but it is
prohibitively expensive. \textit{AutoGeoLabel} can enable rapid label generation for existing petabytes
of data \cite{opentopography}. 
For our test case, point cloud data was collected in 2017, with approx.\ 10 points per
square meter density \cite{nycLiDAR}. A small subset of point cloud data was classified into broad
classes associated with water, ground, and bridges. However the majority of points fell into the
unclassified category. 
The data volume is in excess of $1$ terabyte, and the data is made available as about
$1900$ \texttt{LAS} 3D point cloud files \cite{LAS2019}.

\subsection{Land Cover Data}
\label{sec:LandCoverData}
The LiDAR data of Sec.\ \ref{sec:LiDAR} was further processed by NYC in combination with high resolution,
multi-spectral imagery to generate an $8$--class land cover dataset at $0.5$ meters resolution \cite{nyc-landcover}.
The $8$ classes read: \textit{Tree Canopy}, \textit{Grass/Shrub}, \textit{Bare Soil}, \textit{Water},
\textit{Buildings}, \textit{Roads}, \textit{Other Impervious}, and \textit{Railroads}. Each bears
accuracy in classification above 90\%. The data was also adjusted based on previous city surveys
on roads, rails, and building footprints to further improve classification accuracy. We employ the
\textit{Land Cover} data as a ground truth validation set of \textit{AutoGeoLabel}-generated classes.
Such land cover classes, as obtained for NYC, is not readily available for most parts of the world.
In many cases classification from the \textit{OpenStreetMap} (OSM) project \cite{openstreetmap} is the best at hand.   

\subsection{Geospatial Platforms}
Open-source geospatial data volume exceeds petabytes making it comparable in volume
to data generated by social media \cite{social}. Multiple geospatial platforms exist
where images are stored either as objects \cite{google-ee,aws,planetary} or as indexed pixels ready for computation
\cite{whitby2017geowave,pairs}. IBM PAIRS does index all pixels exploiting a set of predefined
spatial grids to align data layers. This approach renders ideal for search of similarities across
large geographies. Also, the framework proofs efficient to apply similar processing across
multiple datasets.

Current efforts to enable widespread analytics on geospatial data focus on automating machine
learning \cite{autogeo,auto-geos} to lower the user's effort in creating training data,
train models, and aggregate generated results. 
If label data is sparse or missing, \textit{AutoGeoLabel} can quickly generate the required classes.

\section{Methods}
\label{sec:methods}

\subsection{Feature Extraction from Point Cloud Data}
\label{sec:FeatureExtractPointCloud}
Open-source earth observation data libraries such as the \textit{Geospatial Data Abstraction Library}
(GDAL) \cite{gdal} and the \textit{Point Data Abstraction Library} (PDAL) \cite{pdal} enable handling
and manipulation of a plurality of geospatial data formats. Python application programming
interfaces (API) wrapping these libraries provide data scientists low-barrier access to G/PDAL functionalities.
The libraries also offer an easy way to filter and sort information based on a set of statistical attributes.
\textit{AutoGeoLabel} constructs processing pipelines built on GDAL, PDAL, and Python's NumPy module
\cite{2020NumPy-Array}: Raw point cloud data is rasterized as detailed below with the aid of PDAL.
Those rasters get reprojected and aligned into a nested grid utilizing GDAL. Once curated with IBM PAIRS,
fusion with other data layers such as multi-spectral imagery allows queries to pull data cubes
as NumPy arrays through the \textit{PAIRS Python API wrapper} (PAW) \cite{PAW2019} for machine\slash
deep learning consumption (cf.\ e.g. \textit{PyTorch tensors} \cite{paszke2017automatic}).
\begin{figure}
    \centering
    \includegraphics[width=0.95\columnwidth]{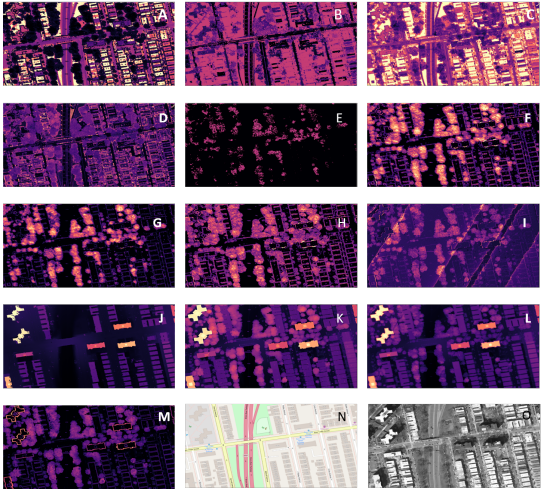}
    \caption{\label{fig:RasterizedStatFeaturesSample}
    A--M: Statistical LiDAR features, cf.\ Tab.\ \ref{tab:PointCloudStats}, as extracted from
    the original point cloud for a sample area in NYC.
    Each layer is stored as individual raster grid with spatial resolution of $\sim0.5$ meters.
    N: depicts OSM labels of the same area as reference; with corresponding orthophoto (O) to the right.}
\end{figure}

\begin{table}[tb]
    \centering
    \caption{Point Cloud Statistics}
    \label{tab:PointCloudStats}
    \begin{tabular}{l c r}
        \hline
        \bf attribute & \bf statistics & \bf Fig.\ \ref{fig:RasterizedStatFeaturesSample} index \\
        \hline
		                     & minimum $r_-$ &  A \\
		reflectance $r$      & maximum $r_+$ &  B \\
		                     & mean $\overline{r}$ & C \\
		                     & standard deviation $r_\Delta$ &  D \\
		\hline
		                     & minimum $c_-$& E \\
		                     & maximum $c_+$& F \\
		count $c$            & mean $\overline{c}$&  G \\
		                     & standard deviation $c_\Delta$&  H \\
		                     & sum $\sum$&  I \\
		\hline
		                     & minimum $e_-$&  J \\
		elevation $e$        & maximum $e_+$&  K \\
		                     & mean $\overline{e}$&  L \\
		                     & standard deviation $e_\Delta$&  M \\
        \hline
    \end{tabular}
\end{table}

For LiDAR data, attributes such as \textit{Intensity}, \textit{Number of Returns}, and \textit{Count} is
recorded to quantify the laser light reflected by the probing pulse sent from the airborne LiDAR device.
While the point cloud may get automatically classified by the data provider, it typically requires specialized,
commercial, and compute intensive software.

Rather than the use of proprietary data processing, in general, \textit{AutoGeoLabel} employs simple
statistical features extracted from high spatial resolution, high quality data. As a demonstration,
here, we present the approach for LiDAR point clouds. The spatial distribution of the point cloud is
determined given a local neighborhood parametrized by the user. For the case presented, an area of linear
extent of approx.\ $2.5$ square meters is defined. It hosts roughly $25$ data points for the evaluation of
statistical features listed in Tab.\ \ref{tab:PointCloudStats}. As the area of aggregations slides
across the coverage area at grid size $0.3$ meters, it determines the features for all data points falling within that area.

In this processing step, the three--dimensional point cloud data is converted to two-dimensional images
where each statistical feature is stored as a separate, curated, and indexed raster layer in IBM PAIRS.
An example of the extracted features for an area in NYC is presented in Fig.\ \ref{fig:RasterizedStatFeaturesSample}.
While the use case presented here exploits simple statistical features like the minimum, maximum,
mean, and standard deviation, higher moments of the distribution like kurtosis,
skewness, etc.\ may get calculated to add further (non-linear) statistical raster layers serving
as additional features. As the point cloud data is converted to $13$ distinct raster layers where
each one stores a statistical feature, visual inspection indicates that vegetation, buildings, roads,
and bare land is the dominant land cover features comprising most of information in the data layers.

\subsection{Rule-Based Labeling from LiDAR surveys}
\label{sec:RuleBasedLabeling}

In order to distill classification information from the rasterized LiDAR statistics generated
in the previous Sec.\ \ref{sec:FeatureExtractPointCloud}, we exploit rules drawn from physical
characteristics when the classification objects get probed by the LiDAR laser pulse:
\begin{itemize}
    \item \textit{buildings}: The firm surface of rooftops is most likely to (partially) reflect the
          laser pulse by a single return. Moreover, compared to the overall size of the building,
          flat roofs bear little variation in elevation. Thus, pseudo-RGB imagery with channels
          encoding minimum $e_-$, maximum $e_+\approx e_-$, and standard deviation $e_\Delta\approx0$
          of elevation measurements will most prominently discriminate buildings in top-down
          airborne LiDAR survey data.
    \item \textit{vegetation}: In contrast to rooftops, vegetation allows for strong variation
          in elevation measurements from LiDAR laser pulses, $e_\Delta\gg0$: As the laser penetrates
          a tree's canopy it might get reflected multiple times by branches and foliage at various
          elevation levels. Moreover, in contrast to a single return $c_-=c_+=1,~c_\Delta=0$ with rooftops,
          multiple pulses will bounce back to the detector, i.e.\ $c_+\gg1$ and $c_\Delta\gg0$
    \item \textit{roads}: Given global terrain slopes have been leveled to zero for elevation
           statistics\footnote{as easily performed in preprocessing of LiDAR point clouds,
           for an application cf.\ e.g.\ \cite{albrecht2019learning}; alternatively,
           fusion with existing elevation models is an option}, $e_-\approx0$. In addition,
           lane markers typically contain reflective particles with mirror like properties
           when illuminated by a laser, $\overline{r}\gg0$. In contrast, the black surface
           of asphalt absorbs a major portion of the laser pulse, $r_-\approx0$.
    \item \textit{water body}: We mention the option of no-data areas in rasterized LiDAR data
          statistics. While the projection of the irregular point cloud onto Earth's surface
          may yield areas of varying point cloud density, larger patches of void area
          typically stem from full absorption of the laser pulse. Water is a prominent land
          class where close-to-zero laser signal is returned, $c_+=0$.
\end{itemize}
Generating pseudo-RGB images $(R,G,B)$, Tab.\ \ref{tab:LabelRules} summarizes
the rules applied to infer classification maps for buildings, roads, and vegetation with
$\langle\cdot\rangle$ spatial averaging of a scene\slash tile. $r_{\max}$ and $e_{\max}$ denote maximum reflectance and maximum
(local, global terrain removed) elevation, respectively. While the thresholding rules for buildings and vegetation
is intrinsically defined based on averaging a given (pseudo-)image patch, establishing rules for road
labeling on laser reflectance values is statically defined. Typically, building height and vegetation
types significantly vary from one geo-location to another. However, in a crude approximation, we consider
constant laser reflectance of e.g.\ road lane markers and road surface ($r_->.1r_{\max}\land\overline{r}<.6r_{\max}$)
on local elevation ground zero ($e_-<.1e_{\max}$) for various geo-locations.

\begin{figure}
    \centering
    \includegraphics[width=0.95\columnwidth]{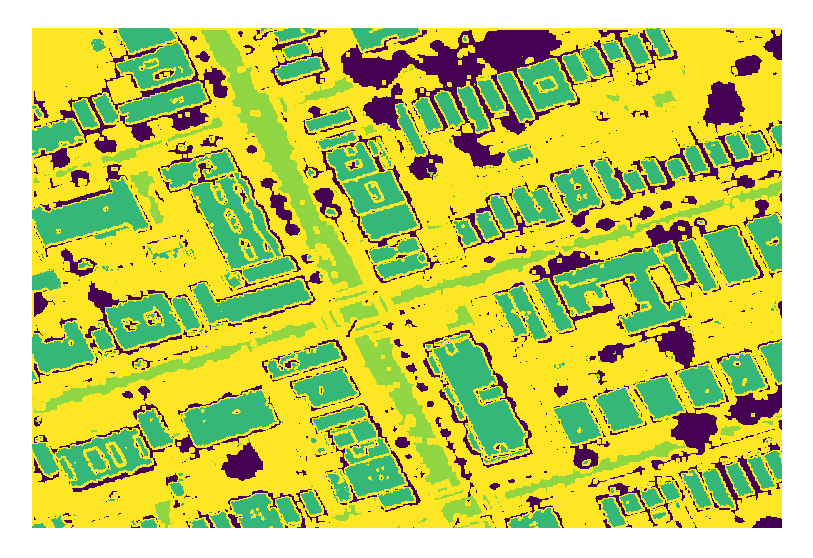}\\
    \includegraphics[width=0.95\columnwidth]{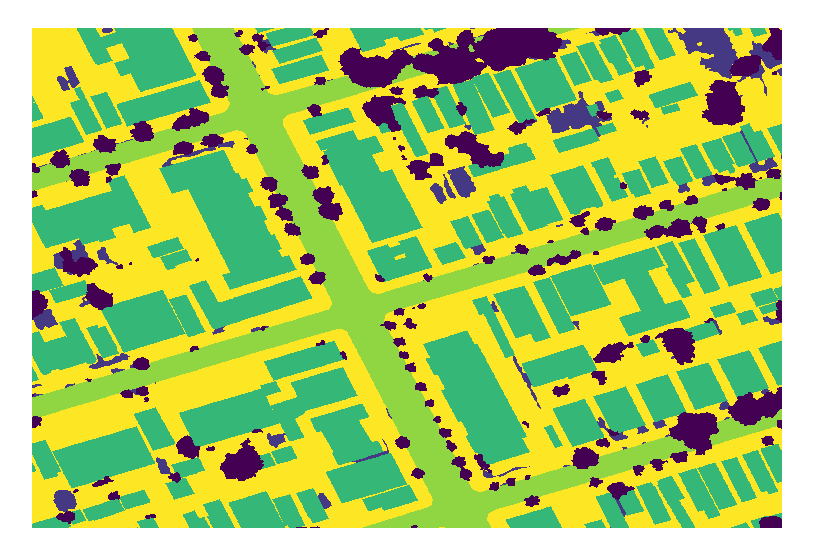}\\
    \caption{Rule-based land cover classification utilizing rasterized LiDAR statistics
    (top, cf.\ Secs.\ \ref{sec:FeatureExtractPointCloud}, \ref{sec:RuleBasedLabeling}) compared to the ground truth
    (bottom, cf.\ Sec.\ \ref{sec:LandCoverData}) exemplified by a block in the neighborhood Corona, Queens, NYC.
    Qualitatively, the classification by LiDAR data statistics reproduces the scene containing buildings, vegetation,
    and roads. Quantitative measures, cf.\ Tab.\ \ref{tab:AccuracyAssessmentRuleBasedLabeling} and Fig.\ \ref{fig:RuleBasedLabelsAccuracyAssessment},
    indicate noise for which its root is detailed in Sec.\ \ref{sec:ValAutoLabel} and Fig.\ \ref{fig:tSNEVisualization}.
    }
    \label{fig:RuleBasedLiDARStatsLabeling}
\end{figure}

The top of Fig.\ \ref{fig:RuleBasedLiDARStatsLabeling} exemplifies the rule-based label generation map.
It is depicted the classes \textit{vegetation} (dark madder purple), \textit{roads} (lime green), \textit{buildings}
(dark green), and \textit{bare land} (yellow). Bare land serves as auxiliary class
for all geo-locations identified as neither road, building, nor vegetation. Fig.\ \ref{fig:RuleBasedLiDARStatsLabeling}
(bottom) plots the ground truth labels of the corresponding area. Apparently, qualitative reconstruction
of the scene is possible.
\begin{table}[tb]
    \centering
    \caption{\label{tab:LabelRules}Labeling Rules from LiDAR Statistics}
    \tiny
    \begin{tabular}{lcl}
        \hline
        \bf class & \bf pseudo (R,G,B) & \bf binary classification rule \\
        \hline
        buildings          & $\left(e_-, e_\Delta, e_+\right)$
        & $e_->\langle e_-\rangle\land e_\Delta<\langle e_\Delta\rangle\land e_+>\langle e_+\rangle$\\
        vegetation         & $\left(c_+, e_\Delta, c_\Delta\right)$
        & $c_+>\langle c_+\rangle\land e_\Delta>\langle e_\Delta\rangle\land c_\Delta>\langle c_\Delta\rangle$\\
        roads              & $\left(r_-, \overline{r}, e_-\right)$
        & $r_->.1r_{\max}\land\overline{r}<.6r_{\max}\land e_-<.1e_{\max}$ \\
        \hline
    \end{tabular}
\end{table}

\subsection{Validation of Automated Label Generation}
\label{sec:ValAutoLabel}

\begin{figure}
    \centering
    \includegraphics[width=0.45\columnwidth]{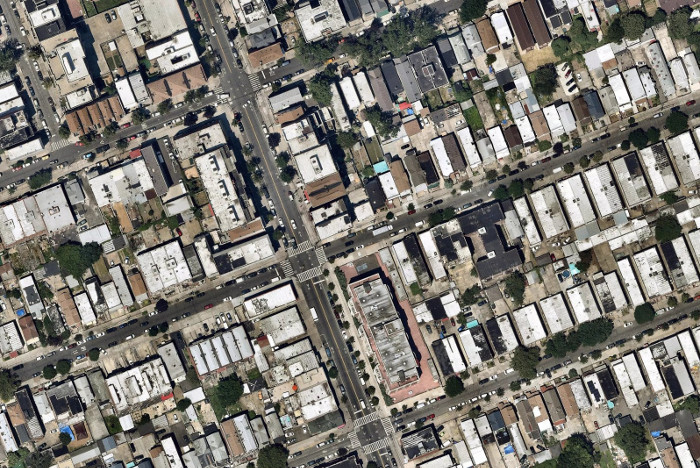}
    \includegraphics[width=0.45\columnwidth]{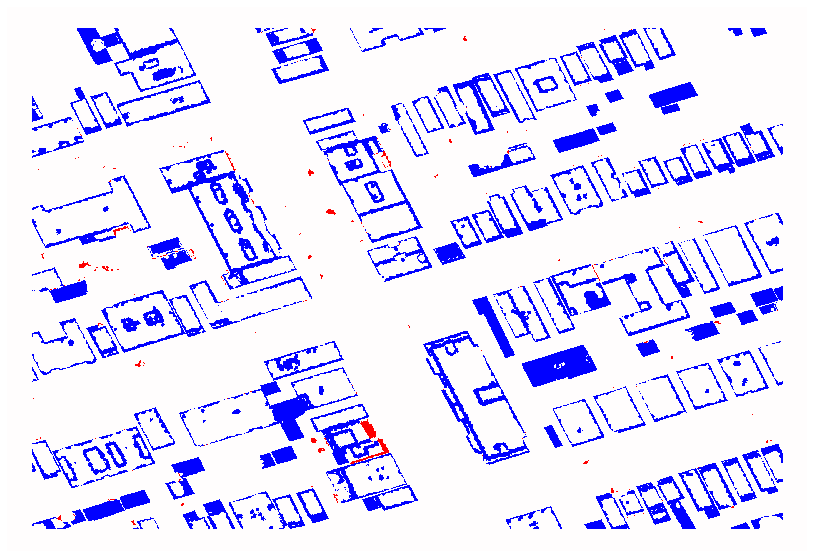}\\
    \includegraphics[width=0.45\columnwidth]{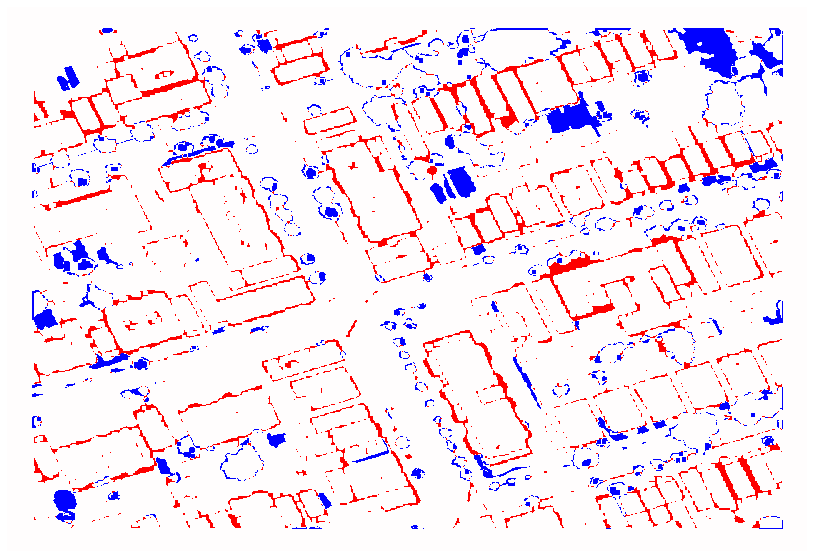}
    \includegraphics[width=0.45\columnwidth]{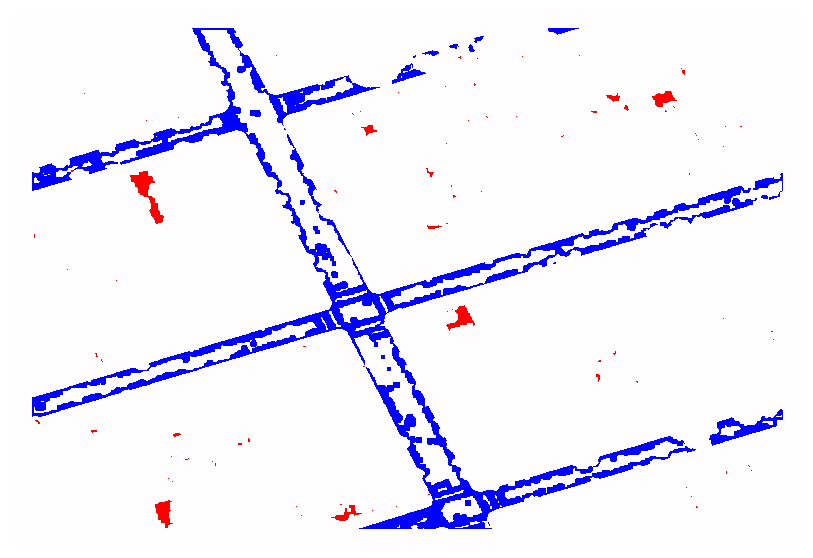}\\
    \caption{Visual accuracy evaluation of rule-based labeling from LiDAR measurements of the urban block exemplified in Fig.\ \ref{fig:RuleBasedLiDARStatsLabeling}. The top left orthophoto provides a visual impression of the scene. The remaining
    sub-figures separately investigate the accuracy of buildings (top right), vegetation (bottom left), and roads (bottom right).
    White color coding marks correct labeling, blue brands missed labels (false negative, $FN$),
    and red distinguishes incorrect class labelling (false positive, $FP$).
    Tab.\ \ref{tab:AccuracyAssessmentRuleBasedLabeling} quantifies the assessment.}
    \label{fig:RuleBasedLabelsAccuracyAssessment}
\end{figure}

A quantitative analysis of accuracy reveals the combination of Fig.\ \ref{fig:RuleBasedLabelsAccuracyAssessment}
and Tab.\ \ref{tab:AccuracyAssessmentRuleBasedLabeling}. In particular, for each class detected, we compute
the binary classification accuracy measures \textit{precision} $P$ and \textit{recall} $R$ according to
\begin{align}
    P=\frac{TP}{TP+FP}
    \quad\text{and}\quad
    R=\frac{TP}{TP+FN}
    \quad.
\end{align}
Given the \textit{Land Cover} ground truth labels (cf. Sec.\ \ref{sec:LandCoverData}) and a class to evaluate
(here: roads, buildings, or vegetation), precision $P$ defines the amount of rule-based labeled pixels
correctly identified (\textit{true positive}: $TP$) in proportion to all pixels labeled as given class
(type I error, \textit{false positive}: $FP$). Similarly, recall $R$ encodes type II errors
(\textit{false negative}: $FN$) by forming the ratio of all true positive pixels relative to all
ground truth pixels of a class. Accordingly, overall (class-specific) accuracy $acc$ is defined by all
the truly classified pixels ($TP+TN$, with \textit{true negative}: $TN$) in proportion to the total
number of pixels:
\begin{equation}
    acc=\frac{TP+TN}{TP+TN+FP+FN}\quad.
\end{equation}

As visually depicted in Fig.\ \ref{fig:RuleBasedLabelsAccuracyAssessment}, identification of buildings and roads is
dominated by false negatives (blue). Reversely, false positives (red) govern identification of vegetation.
For labeling of buildings and vegetation such inverse relation of $FP$ and $FN$ roots in the physical reflectance
properties of the LiDAR measurements at the edge of buildings: As the laser partially hits the outline of a building,
multiple pulses get reflected back to the detector from the rooftop, the building's face, and the ground---a scenario
the rule-based labeling for vegetation in Tab.\ \ref{tab:LabelRules} is sensitive to. In fact, this source of label
noise might get exploited for building footprint extraction utilizing traditional computer vision (post-)processing
steps such as morphological filtering \cite{sigmund2007morphology} or a straight line detector \cite{kiryati1991probabilistic}.

Except for the building edge labeling artefact, low false positive/negative rates on bulk objects such as buildings and
vegetation allow for a relatively high $F_1$ score determined by the harmonic mean of precision and recall:
\begin{equation}
F_1 = \frac{1}{\frac{1}{2}(1/P + 1/R)}=\frac{2PR}{P+R}
\quad.
\end{equation}
Nevertheless, a fourth quantity, the \textit{Intersection over Union} ($IoU$) is required to complement the accuracy
evaluation:
\begin{equation}
IoU = \frac{\vert C\cap T\vert}{\vert C\cup T\vert}
\quad,
\end{equation}
with $C$ the set of rule-based labels of a given class, and $T$ the corresponding ground truth.
$\vert\cdot\vert$ denotes the set size operator which returns geospatial area covered.
This way, it is measured the degree of spatial misalignment of rule-based labels wrt.\ the ground truth.
Since label noise dominates at the boundary of classification objects, $IoU$ scores low for buildings
and vegetation. Results for road labeling suffer from covering vegetation and \textit{curb noise} such
as parked vehicles, power lines, traffic lights, light poles, etc.

However, as Fig.\ \ref{fig:RuleBasedLiDARStatsLabeling} visually demonstrates, the bulk of objects gets correctly
labeled by the rule-based approach. A fact imprinted in the $acc$-measure for each class.
Hence, we propose a challenge to the large-scale data mining remote sensing community: employ
rule-based rasterized LiDAR statistics as automatically generated, noisy labels to benchmark weakly
supervised semantic segmentation methodologies. In particular, the NAIP orthoimages
\cite{naip-orthophoto} in combination with the NYC LiDAR data \cite{nycLiDAR} provides model input.
\begin{table}[tb]
    \centering
    \caption{Accuracy Assessment of rule-based Labeling}
    \label{tab:AccuracyAssessmentRuleBasedLabeling}
    \scalebox{.9}{
      \begin{tabular}{l|ccccc}
         \hline
         \bf class & \bf precision $P$ & \bf recall $R$ & \bf $F_1$-score &\bf acc  & \bf Intersection \\
                   &                   &                &                 &         & \bf over Union $IoU$ \\
         \hline
         buildings  & .98 & .62 & .76 & .88 & .61\\
         vegetation & .52 & .60 & .55 & .90 & .38\\
         roads      & .91 & .44 & .59 & .93 & .42\\
         \hline
      \end{tabular}
    }
\end{table}

\subsection{LiDAR Statistics Clustering for Label Generation}

\begin{figure}
    \centering
    \includegraphics[width=0.95\columnwidth]{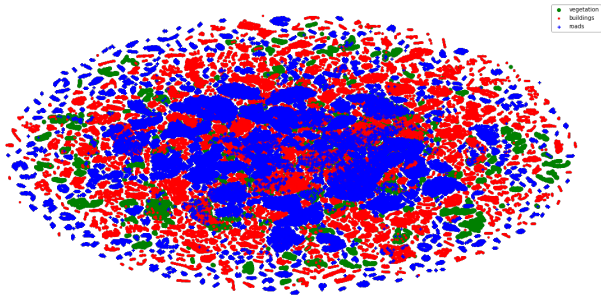}\\
    \caption{t-SNE embedding of the $13$--dimensional LiDAR raster statistics, cf.\ Tab.\ \ref{tab:PointCloudStats}.
    Randomly picked land classification samples of \textit{vegetation} (green, $\boldsymbol{\cdot}$), \textit{buildings}
    (red, $\cdot$), and \textit{roads} (blue, $+$) is plotted. It becomes apparent that near to perfect class
    separation requires a strongly non-linear function such as e.g.\ modelled by artificial neural networks.}
    \label{fig:tSNEVisualization}
\end{figure}

To further investigate the label noise, we visualize data clustering in the multi--dimensional space of
rasterized LiDAR statistics. Specifically, we utilized the $13$ raw statistics features $X=(A,B,\dots,M)$
of Tab.\ \ref{tab:PointCloudStats} to non-linearly project these into two dimensions for plotting through
\textit{t-Distributed Stochastic Neighbor Embedding} (t-SNE) \cite{vandermaaten08a}: $(x,y)=tSNE(X)$.
Random sampling of geo-locations generates a ground-truth $l\in\{\text{building},\text{road},\text{vegetation}\}$
labeled set of data points $\{(X_1,l_1),(X_2,l_2),(X_3,l_3),\dots\}$. The corresponding t-SNE projection
$\{(x_1, y_1),(x_2, y_2),(x_3,y_3),\dots\}$ is presented in Fig.\ \ref{fig:tSNEVisualization} with color
coding $\{l_1,l_2,l_3,\dots\}$.

While cluster sizes and distances between strongly depend on t-SNE initial conditions and
parameter settings, qualitative cluster structure in high-dimensional space may get read off from the
two-dimensional embedding. In particular, we ran t-SNE experiments with various settings and differing data samplings.
Repeatedly, it yielded qualitatively similar results as exemplified by Fig.\ \ref{fig:tSNEVisualization}:
The three classes associated with vegetation, buildings, and roads are well separated. Thus, distinct sets of
classes can be defined through a combination of LiDAR statistics layers in order to identify those.
However, standard clustering methods like $k$-means---even when adapted to variable number of clusters
\cite{hamerly2004learning}---is likely to fail due to the highly nonlinear separation of classes: As
apparent from Fig.\ \ref{fig:tSNEVisualization}, a single class like vegetation is associated with a
number of clusters well separated.

Highly non-linear functions can get modelled by artificial neural networks. Research in self-supervised
learning recently demonstrated the generation of expressive feature vectors without the need for any
labels \cite{jing2020self}. Performance of self-supervised learning is typically measured by the accuracy
of \textit{downstream tasks} that employ the learnt feature representation $z$. Specifically, object classification
is evaluated by training a single-layer of fully connected neurons attached to $z$ on a small set of
ground truth labels. Hence, a future research direction in \textit{AutoGeoLabel} may exploit self-supervised
learning to improve upon the noisy, rule-based label generation process with the aid of a small set of ground
truth labels.
 
\section{Application}

We demonstrate an industrial application of labels generated by \textit{AutoGeoLabel} to identify trees and quantify carbon sequestration \cite{klein2019,carbon}.
LiDAR statistics can identify tree's location, extract canopy diameter and calculate the total carbon sequestered in trees.
\textit{AutoGeoLabel}-generated vegetation labels are converted to polygons. Subsequently, the mask is used to crop the
maximum elevation LiDAR statistics data, $e_+$. The obtained image with vegetation height is segmented using a watershed
method to delineate the tree crown diameter \cite{watershed}. Then, the tree crown polygons can be used as
another simple filter layer where the eccentricity \cite{eccentricity} of polygons helps to eliminate features that
appears elongated to preserve rounded\slash circular features, only. In addition, filtering is employed to remove features
that are too small or too large to be associated with a tree. 

Manually labeled tree species data is used to create a classifier associating tree species labels \cite{treespecies}. Four dominant tree species are used to reclassify all trees within NYC \cite{carbon}.
Tree height is extracted from the LiDAR height where the height is adjusted taking into account
the ground elevation---resulting in absolute tree height.
\begin{figure}
    \centering
    \includegraphics[width=0.95\columnwidth]{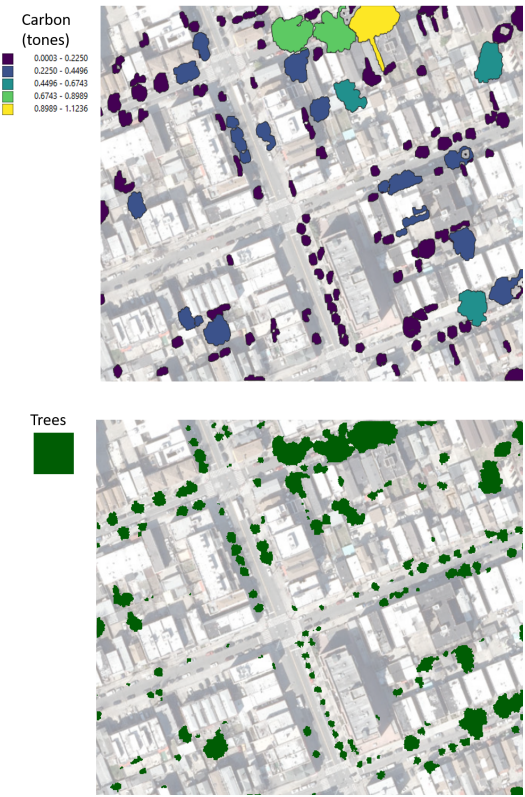}
    \caption{Application: Total carbon stored in trees (top) calculated based on tree species and tree dimensions
    derived from LiDAR measurements for geospatial sample area of Fig.\ \ref{fig:RuleBasedLiDARStatsLabeling}.
    The corresponding ground truth image (bottom) depicts the distribution of the urban forest, cf. Fig.\ \ref{fig:RuleBasedLiDARStatsLabeling} (bottom).}
    \label{fig:CarbonStorage}
\end{figure}
Quantification of carbon stored in trees follows a procedure outlined before \cite{carbon}.
The method presented here is different from previous works as it exclusively relies on LiDAR data
to identify vegetation using automatically created labels, delineate tree canopy, and calculate
tree heights. The only additional data required is tree species information \cite{treespecies, carbon}.
Tree identification and carbon sequestration offer a way for city management to
better plan tree replanting, and to efficiently quantify total carbon stored in urban forest.

\section{Conclusion \& Perspectives}

In this paper we presented \textit{AutoGeoLabel}; a framework to address automatic data labeling for
geospatial applications. Many industrial and scientific solutions can benefit from automating label
generation to overcome the challenge of manual image
annotation---a labor- and time-consuming effort. Based on an airborne LiDAR survey for New York City, 
we demonstrated and explored a novel approach to use simple statistical features of remote sensing
data in order to create data classes. Weakly supervised,
and self-supervised learning has been argued as promising approaches. Utilizing automatically
generated labels for vegetation, we demonstrated an application to quantify carbon sequestration in
urban forests with no need for explicit tree segmentation from e.g. orthophotos. 

In perspective of industry applications, there is great potential of \textit{AutoGeoLabel} to contribute.
E.g. for emergency response during a natural disaster event, many of the existing labeled data acquired under
\textit{normal conditions} may not hold representative of what it is observed on the ground. In such extreme
situations, new labeled data may need to be generated on the fly. One such example is recognizing flooded areas utilizing aerial, or drone images where training data may not exist. Detecting the impact of extreme weather can drive rescue
missions, where assessment of \textit{change} from normal conditions like water extent and potential damage
(estimates of depth of water) require creation of data features that can be used to delineate the boundary
of overflown water. Then, information related to the number of flooded houses and roads may help to coordinate the best routing for rescue missions.

{
\small
\bibliographystyle{IEEEtran}
\bibliography{Paper_IEEEBigData2021_AlbrechtMariannoKlein_AutoGeoLabel}
}
\end{document}